\documentclass[a4paper]{aastex}
\usepackage{spr-astr-addons} 

\usepackage{url}
\usepackage{times}
\usepackage{mathptmx}  

\def\elem#1[#2]{{}$^{\scriptsize #2}${\rm #1}}
\newcommand{\etal}{{et al.}}
\newcommand{\posc}{{\sc posc}}

\def\muHz{\, \mu{\rm Hz}}
\def\xin{x_{\rm in}}

\def\xout{x_{\rm out}}
\def\yin{\vec{y}_{\rm in}}
\def\yout{\vec{y}_{\rm out}}
\def\CA{{\cal A}}

\def\diff{{\rm d}}
\def\devfp#1#2{\displaystyle {\partial #1 \over \partial #2}}
\def\devfd#1#2{\displaystyle {{\rm d} #1 \over {\rm d} #2}}

\def\im{{\rm i}}
\def\dis{\displaystyle}
\def\chiu{\hbox{\raise2pt\hbox{$\chi$}}}

\begin{document}

\title{Porto Oscillation Code (\posc)}

\author{M\'ario~J.~P.~F.~G.~Monteiro}
\shortauthors{Monteiro}

\affil{Centro de Astrof\'{\i}sica da Universidade do Porto, Rua das Estrelas, 4150-762 Porto, Portugal}
\affil{Departamento de Matem\'atica Aplicada da Faculdade de Ci\^encias,
           Universidade do Porto, Portugal}

\email{mario.monteiro@astro.up.pt}

\begin{abstract}

The {\em Porto Oscillation Code} (\posc) has been developed in 1995 and improved over the years, with the main goal of calculating linear adiabatic oscillations for models of solar-type stars.
It has also been used to estimate the frequencies and eigenfunctions of stars from the pre-main sequence up to the sub-giant phase, having a mass between 0.8 and 4 solar masses.

The code solves the linearised perturbation equations of adiabatic pulsations for an equilibrium model using a second order numerical integration method.
The possibility of using Richardson extrapolation is implemented.
Several options for the surface boundary condition can be used.
In this work we briefly review the key ingredients of the calculations, namely the equations, the numerical scheme and the output.

\end{abstract}


\section{Introduction}\label{sec:intro}

The {\em Porto Oscillation Code} (\posc) was initially developed in 1995 to obtain the frequencies of solar models and envelopes.
The first description of the code has been given in \cite{monteiro96a}.

The objective of this paper is to present a summary on how \posc\ calculates the frequencies of oscillations for stellar models.
The paper starts with the basic linear equations describing the oscillations and how these are formulated to be solved numerically.
The boundary conditions used and their implementation
are also discussed as well as the accuracy of the calculations.
We end by listing some of the output values provided by the code and some of the applications where the results of the code have been used.

\section{Basic equations for linear perturbations}\label{sec:1}

Our objective here is to review the necessary equations for {\em non-radial
adiabatic oscillations} of spherically symmetric non-rotating stars.
By following the work by \cite{unno89} it is possible to start from the hydrodynamic equations (continuity, Poisson and conservation of momentum equations), in order to obtain a set of equations describing the radial dependence of the amplitude functions for small perturbations.
These perturbations correspond to; $P$ for pressure, $\Phi$ for gravitational potential, while $\vec{\xi}$
is the displacement.
The solutions are writen as
\begin{equation}\begin{array}{l}
\dis P(t{,}r{,}\theta{,}\phi) = P_0(r) +
   \tilde P(r)\;\; Y_l^m(\theta{,}\phi)\;
	e^{\im\omega t} \;, \\[5pt]
\dis\Phi(t{,}r{,}\theta{,}\phi) = \Phi_0(r) +
   \tilde\Phi(r)\;\; Y_l^m(\theta{,}\phi)\;
	e^{\im\omega t} \;, \\[5pt]
\dis\vec{\xi}(t{,}r{,}\theta{,}\phi) =
	\left[\xi_r(r),\; \xi_h(r)\; {\partial Y_l^m\over \partial \theta}, 
	{\xi_h(r) \over \sin\theta}\; {\partial Y_l^m\over \partial\phi} \right]\;
	 e^{\im\omega t} .
\end{array}\end{equation}
Where the equilibrium configuration of the stars is described by the functions; $\rho_0$ (for density), $P_0$ and $\Phi_0$.
Here $t$ is time, $\omega$ the frequency for the oscilating solutions, $(\theta,\phi)$ the horizontal variables while $r$ is radial distance and $Y_l^m(\theta{,}\phi)$ the spherical harmonics characterized by the integer numbers $l$ (mode degree) and $m$ (azimutal order with $m{=}{-}l,..,0,..,l$).

By considering an equation for adiabatic perturbations and after eliminating the dependence on the horizontal coordinates and time, the equations describing the radial amplitude of the small perturbations are obtained in the following form;
\begin{equation}\begin{array}{l}
\dis \left(1 {-} {S_l^2 \over \omega^2} \right) {\tilde P \over \rho_0} -
	{1 \over r^2} \left( g_0 {-} c_o^2\; {{\rm d} \over {\rm d} r} \right) 
	(r^2\xi_r) + {S_l^2 \over \omega^2} \tilde\Phi = 0,\!\!\! \\[15pt]
\dis {1 \over \rho_0} \left({g_0 \over c_0^2} + {\diff \over \diff r} \right)\;
	\tilde P - (\omega^2{-}N_0^2)\; \xi_r - \devfd{\tilde\Phi}{r} = 0 \;, \\[15pt]
\dis \tilde P + {\rho_0 c_0^2 N_0^2 \over g_0}\; \xi_r -
	{S_l^2 \over 4\pi G} \tilde\Phi +
	{c_0^2\over 4\pi Gr^2}\;
	{\diff \over \diff r} \left(r^2\devfd{\tilde\Phi}{r}\right) = 0.
\end{array}\label{eq:basic}\end{equation}

The equilibrium structure in these equations is also characterized by quantities as gravity $g_0$ and sound speed $c_0$;
\begin{equation}\begin{array}{l}
\dis g_0 = - \devfd{\Phi_0}{r} = - {1 \over \rho_0}\; \devfd{P_0}{r} \;,\\[15pt]
\dis c_0^2 = {\Gamma_{1{,}0}\; P_0 \over \rho_0} \qquad
{\rm with}\quad \Gamma_1 \equiv \left( \devfp{\log P}{\log\rho} \right)_S  \;,
\end{array}\end{equation}
where the derivate has been calculated at fixed entropy $S$.
There are also two characteristic frequencies; the {\em Lamb frequency} $S_l$ and the {\em buoyancy frequency} $N_0$ (also known as the Brunt-V\"aiss\"al\"a frequency), corresponding to
\begin{equation}\begin{array}{l}
\dis S_l^2 = l(l{+}1)\; {c_0^2 \over r^2} \;, \\[15pt]
\dis N^2_0 = g_0 \left( {1 \over \Gamma_{1{,}0}}\; \devfd{\log P_0}{r} - 
	\devfd{\log\rho_0}{r} \right) \cr
\dis \qquad 	= - g_0 \left( {g_0\rho_0 \over \Gamma_{1{,}0}\; P_0} + 
	\devfd{\log\rho_0}{r} \right) \;.
\end{array}\end{equation}

If we consider the following dimensionless variables
\begin{equation}\begin{array}{l}
\dis y_1 = {\xi_r \over r} \\
\dis y_2 = {\omega^2 \over g}\; \xi_h
	= {1 \over rg}\; \left( {\tilde P \over \rho} - \tilde\Phi\right) \\
\dis y_3 = {\tilde\Phi \over rg} \\
\dis y_4 = {1 \over g}\; \devfd{\tilde\Phi}{r} \;,
\end{array}\end{equation}
the equations can be written as
\begin{equation}\begin{array}{l}
\dis r\; \devfd{y_1}{r} = \left( {rg_0 \over c_0^2} {-} 3 \right) y_1 +
	{rg_0 \over c_0^2}\; \left( {S_l^2 \over \omega^2} {-} 1 \right)
	y_2 - {rg_0 \over c_0^2}\; y_3 \cr
\dis r\; \devfd{y_2}{r} = - {r \over g_0}\; (N_0^2{-}\omega^2) y_1 -
	\Big( {rg_0 \over c_0^2}+\devfd{\log \rho_0}{\log r}+\cr
	\dis \qquad +4\pi G\;
	{r\rho_0 \over g_0} - 1\Big) y_2 - \left({rg_0 \over c_0^2} + \devfd{\log\rho_0}{\log r}\right) y_3 \cr
\dis r\; \devfd{y_3}{r} = \left( 1 - {4\pi Gr\rho_0 \over g_0} \right) y_3 + y_4 \\
\dis r\; \devfd{y_4}{r} = - {4\pi Gr^2\rho_0 \over c_0^2}\;
	{c_0^2N_0^2 \over g_0^2}\; y_1 -
	{4\pi Gr^2\rho_0 \over c_0^2}\; y_2 + \cr
\dis\qquad 	\left[ l(l{+}1) - {4\pi Gr^2\rho_0 \over c_0^2}\right] y_3 - {4\pi Gr^2\rho_0 \over c_0^2}\; {c_0^2 \over rg_0}\; y_4 \;.
\end{array}\end{equation}

These form the set of equations we need to solve to obtain the radial behaviour of linear adiabatic oscillations of spherically symmetric stars.

\section{The equilibrium model}

In order to describe the reference/equilibrium model we consider the following dimensionless functions of the equilibrium structure
(as defined by \citealt{adipls})
\begin{equation}\begin{array}{l}
\dis x \equiv {r \over R} \cr
\dis\quad a_1 \equiv {m_{r{,}0} \over r^3}\; {R^3 \over M} \\
\dis \quad a_2 \equiv - {1 \over \Gamma_{1{,}0}}\; \devfd{\log P_0}{\log r} =
	{rg_0 \over c_0^2} \\
\dis \quad a_3 \equiv \Gamma_{1{,}0} \\
\dis \quad a_4 \equiv {1 \over \Gamma_{1{,}0}}\; \devfd{\log P_0}{\log r} -
	\devfd{\log\rho_0}{\log r} = {r \over g_0}\; N_0^2 \\
\dis \quad a_5 \equiv {4\pi r^3\rho_0 \over m_{r{,}0}} \;,
\end{array}\label{eq:apar}\end{equation}
where $M$ and $R$ are respectively the total mass and radius of the star, while $m_r$ is the mass within a sphere of radius $r$.
These 5 functions are the result of an evolution code, being necessary as the input of the oscillation code.

The four first order differential equations for small amplitudes can now be written simply as
\begin{equation}\begin{array}{l}
\dis x\; \devfd{y_1}{x} = (a_2{-}3) y_1 + \left[ \dis {l(l{+}1) \over \sigma^2}\; a_1 -
	a_2 \right] y_2 + a_2 y_3 \\[10pt]
\dis x\; \devfd{y_2}{x} = \left( {\sigma^2 \over a_1} - a_4 \right)
	y_1 + (1{+}a_4{-}a_5) y_2 - a_4 y_3 \\[10pt]
\dis x\; \devfd{y_3}{x} = (1{-}a_5) y_3 + y_4 \\[10pt]
\dis x\; \devfd{y_4}{x} = a_4a_5 y_1 + a_2a_5 y_2 +\left[ l(l{+}1) - a_2 a_5
	\right] y_3 - \\
	\qquad\qquad  -a_5 y_4 \;,
\end{array}\end{equation}
where we have introduced the reduced frequency
\begin{equation}
\sigma^2 = {R^3 \over GM}\;\; \omega^2 \;.
\end{equation}

We may write these equations in a vectorial form by defining the matrix, with $L^2{=}l(l{+}1)$,
\begin{equation}
{\cal A} = \left[ \begin{matrix}
       a_2{-}3 & {\dis L^2 \over \dis \sigma^2}\,
			a_1{-}a_2 & a_2 & 0 \cr
		{\dis \sigma^2 \over \dis a_1} {-} a_4 &
			a_4{-}a_5{+}1 & - a_4 & 0 \cr
		0 & 0 & 1{-}a_5 & 1 \cr
		a_4a_5 & a_2a_5 & L^2{-}a_2a_5 & - a_5 \end{matrix} \right] \;.
\label{eq:matrixa}
\end{equation}
The system of differential equations is then simply written as
\begin{equation}
x\; \devfd{\vec{y}}{x} = {\cal A} \cdot \vec{y} \;,
\label{eq:difosc}
\end{equation}
where the vector $\vec{y}$ has the components $(y_1,y_2,y_3,y_4)$.

\section{Boundary conditions}

To complete the required equations it is also necessary to define four boundary conditions.
The solution is to be found by integrating the equations between the
centre of the star ($r{=}0$ or $x{=}0$) and the top of the atmosphere ($r{\ge}R$ or $x{\ge}1$).
So, in fact we shall be establishing two boundary conditions at $x{=}0$ and other two at the surface.
The result is an eigenvalue problem with solutions existing for discrete
values of $\sigma$.
These are the eigenvalues associated to the corresponding eigenfunctions, that must satisfy the boundary conditions.

\subsection{At the centre}

Since we have four dependent variables, the interior boundary conditions
correspond to fix the values for two of the dependent variables.
The other two are then related to these.

The boundary conditions have to guarantee that the solutions are
regular in the singular point, $x{=}0$, of the differential equations.
So, we start by determining the limiting behaviour of the $a_i$'s
when $x{\rightarrow}0$.
Considering that for $x{\equiv}r{/}R{\ll}1$ we can write (the subscript ``c'' stands for the value at $x{=}0$);
\begin{equation}
\rho \sim \rho_c \;,\quad
m_r \sim {4\pi R^3 \over 3}\; \rho_c\; x^3 \;, \quad{\rm and}\quad
P \sim P_c \;.
\end{equation}

These expressions determine the behaviour of
the $a_i$'s near the centre as follows from the definitions~(\ref{eq:apar});
\begin{equation}
a_1 \sim {4\pi \over 3}\; \bar\rho_c \;,\;
a_2 \sim 0 \;,\;
a_3 \sim \Gamma_{1c} \;,\;
a_4 \sim 0 \;,\;
a_5 \sim 3\; ,
\end{equation}
where $\bar\rho_c{=}R^3\rho_c/M$.

If we now replace these in the definition~(\ref{eq:matrixa})
of the matrix ${\cal A}$, the problem is reduced to a simple
system of differential equations with constant coefficients.
This is,
\begin{equation}
x\; \devfd{\vec{y}}{x}
	\simeq {\cal A}_{\rm c} \cdot \vec{y} \;,
\label{eq:bc-in}
\end{equation}
with
\begin{equation}
{\cal A}_{\rm c} = \left[ \begin{matrix}
   -3 & \dis {4\pi \bar\rho_c \over 3}\;
   {L^2 \over \sigma^2} & 0 & 0 \cr
   \dis {3 \over 4\pi\bar\rho_c}\; \sigma^2 & -2 & 0 & 0 \cr
   0 & 0 & -2 & 1 \cr
   0 & 0 & L^2 & -3 \cr
\end{matrix} \right] \;.
\end{equation}
This has a general solution (non-zero) given by
\begin{equation}
y_j = x^{l{-}2}\; \sum_{i{=}0}^{\infty} {\cal Y}_{ij}\; x^{2i}
	\qquad ;\; j{=}1,2,3,4 \;,
\end{equation}
with
\begin{equation}
{\cal Y}_{01} = {4\pi \bar\rho_c \over 3}\; {l \over \sigma^2}\;\;
	{\cal Y}_{02}
\qquad{\rm and}\qquad
{\cal Y}_{04} = l\;\; {\cal Y}_{03} \;.
\label{eq:bc-cen}
\end{equation}

These are the two boundary conditions at the centre:
from the values of ${\cal Y}_{02}$ and ${\cal Y}_{03}$ it is possible to determine ${\cal Y}_{01}$ and ${\cal Y}_{04}$.

\subsection{At the atmosphere}

Two more boundary conditions need to be imposed at the top of the atmosphere.
In a similar fashion to what has been done for the centre, we now need to established what is the
limiting behaviour for the $a$'s (the subscript ``$S$'' represents in the following the value
at the top of the atmosphere, located at $r_{\rm s}/R{\ge}1$) for $x{\rightarrow}x_{\rm s}$.

There are different options for imposing a boundary condition at the top of the model (surface).
The most commonly used one is to assume an isothermal atmosphere for which we have that,
\begin{equation}
a_1 \sim 1 \;,\;
a_2 \sim a_{2s} \;,\;
a_3 \sim \Gamma_{1s} \;,\;
a_4 \sim a_{2s} \;,\;
a_5 \sim 0 \;.
\label{eq:ai-surf}
\end{equation}
In such an isothermal atmosphere the density decreases exponentially
with radius.
This behaviour allows one to approximate the actual value of $a_{\rm 5s}$ by zero.

The set of equations is now written as,
\begin{equation}
x\; \devfd{\vec{y}}{x}
	\simeq {\cal A}_{\rm s} \cdot \vec{y} \;,
\label{eq:bc-out}
\end{equation}
where the matrix ${\cal A}$, has been approximated using the approximate values of $a_i$ (see \ref{eq:ai-surf}):
\begin{equation}
{\cal A}_{\rm s} = \left[ \begin{array}{lllll}
         a_{\rm 2s}{-}3 & {\dis L^2 \over \dis \sigma^2}\,
		{-}a_{\rm 2s} & a_{\rm 2s} & 0 \cr
		\sigma^2 {-} a_{\rm 4s} &
		1 {+} a_{\rm 4s} & - a_{\rm 4s} & 0 \cr
		0 & 0 & 1 & 1 \cr
		0 & 0 & L^2 & 0 \end{array} \right] \;.
\end{equation}
By redoing the analysis presented in the previous subsection, and using
\begin{equation}
y_j = x^{{-}l}\; \sum_{i{=}0}^{\infty} {\cal Y}_{{\rm s},ij}\; x^{2i}
	\qquad ;\; j{=}1,2,3,4 \;,
\end{equation}
it follows that
\begin{equation}\begin{array}{l}
\dis {\cal Y}_{S,01} = 2\; {\left( L^2/\sigma^2 - a_{2S} \right) {\cal Y}_{S,02}
	+ a_{2S} {\cal Y}_{S,03} \over a_{4S}{+}4{-} \gamma^{1/2}} \;,\\
\dis\qquad \gamma = \left( a_{2S}{-}a_{4S}{-}4 \right)^2 + 4
      (\sigma^2{-}a_{4S}) \left( {L^2 \over \sigma^2}{-}a_{2S} \right) \;, \\[10pt]
\dis {\cal Y}_{S,04} = - (l{+}1)\; {\cal Y}_{S,03} \;. \cr
\end{array}\label{eq:bc-surf1}\end{equation}

When using the first expression one must be careful since the actual
value for $\gamma^{1/2}$ can be imaginary.
If it happens the solution will have a propagating component
at the boundary, which implies that the wave will be loosing
energy at this boundary.
This does not correspond to the type of solutions we are looking for (standing waves).
Therefore we only consider eigenvalues that are real, corresponding to
standing waves, i.e. solutions that are evanescent at the boundaries.

Note that this imposes restrictions on the
values the frequency $\sigma$ can have for possible modes of oscillation.
Solutions are only calculated for $\gamma\ge0$.

Other options for surface boundary conditions are possible (and have been implemented in \posc).
The simplest option is to impose full reflection at the top of the model.
Such a condition is achieved by setting $\delta P{=}0$ at $x{=}x_s$, giving that
\begin{equation}
{\cal Y}_{S,01} = {\cal Y}_{S,02} + {\cal Y}_{S,03} \;,
\label{eq:bc-surf2}
\end{equation}
instead of the first expression in Eqs~(\ref{eq:bc-surf1}).

\section{Calculation of the solutions}

In order to calculated the eigenvalues (frequencies of oscillation)
of a solar model \posc\ uses a simple numerical scheme to solve the set of equations~(\ref{eq:difosc}) with the boundary conditions~(\ref{eq:bc-cen}) and (\ref{eq:bc-surf1}) (or one of the other alternatives).
In this Section we describe briefly how this is done.

The actual expressions implemented in the code are extracted from the basic dimensionless system of 4 differential equations;
\begin{equation}
x\; \devfd{\vec{y}}{x} = \CA \cdot \vec{y} \;,
\label{eq:def-eq}
\end{equation}
where the matrix $\CA$ is given in Eq.~(\ref{eq:matrixa}).

The functions $a_i$ are as listed in Eq.~(\ref{eq:apar}) and known on a mesh from the equilibrium model obtained from solving the stellar structure equations.
We use the dimensionless frequency $\sigma$, related to the actual frequency of oscillation ($\omega$).
As discussed above, under the selected boundary conditions, solutions exist
only for discrete values of $\omega=\omega_{ln}$.
The {\em mode order} $n$ is associated with the radial structure of the different eigenfunctions that exist for the same mode degree $l$ (the code considers spherical stars, and so the solutions are independent of the azimutal order $m$).

These values and the corresponding solution $\vec{y}$ are what we are
trying to find.
The method we use consists in, given a value of the degree $l$,
to determine the values(s) of $\sigma$ that give a continuous solution
at some meeting point (defined below as $x_f$).
This point is where we stop the integration up from the centre,
and the integration down from the atmosphere.
In other words we find the value(s) of $\sigma$ that have a global
solution satisfying all our four boundary conditions.
So what we do in fact is to iterate in $\sigma$ in order to find the values
that give the zeros of a function measuring the fitting of outer
and inner solutions at $x_f$.

\subsection {Numerical variables}
Due to numerical control of errors and precision of the calculation we 
redefine the variables for different regions of the model.
We do so by estimating which regions of the star are evanescent for a given frequency.
The two points used here to define these regions are $\xin$ and $\xout$.
These depend on the model and the values of $\omega$ and degree $l$, corresponding to the roots of the following equation,
\begin{equation}
\omega^2\; (\omega^2 {-} \omega_c^2) - S_l^2\; (\omega^2 {-} N^2) = 0 \;.
\end{equation}
The acoustic cutoff frequency $\omega_c$ used here is determined by
\begin{equation}
\omega_c \equiv {c_0^2 \over 4 H_\rho^2} \left( 1 {-} 2 \devfd{H_\rho}{r}\right) \quad
{\rm where}\quad
H_\rho \equiv \left| \devfd{\log\rho}{r}\right|^{{-}1} \;.
\end{equation}

For the inner (near the centre) evanescent region we redefine the variables according to
\begin{equation}
\yin = \left({x \over \xin}\right)^{2{-}l} \vec{y} \;.
\end{equation}
Here, $\xin$ is the transition point separating this inner region from
the zone where the default variables, as given in Eq.~(\ref{eq:def-eq}), are used. 

For the outer evanescent region (surface layers), above $x{=}\xout$, we use instead
\begin{equation}
\yout = \left({x \over \xout}\right)^{l} \vec{y} \;.
\end{equation}

The equations are integrated from $x{=}0$ to $x{=}\xin$ determining
$\yin$.
From there to a fitting point $x_f$ (well within the oscillatory region)
we calculate the solution using the equations for $\vec{y}$.
Note that the transition from one region to the other is quite natural
considering our definitions $\yin$ and $\yout$ of $\vec{y}$.
On the other hand we integrate inward from $x{=}x_{\rm s}$ to $x{=}\xout$ using instead the equations for $\yout$.
From there, down to $x_f$ we take again the equations for $\vec{y}$.

Resulting from these two integrations we have the two sets of values
at $x{=}x_f$ which are then continuous (after normalization).
Since the system of equations is linear, this is so if and only if the value of $\sigma$ is an eigenvalue.
At this point what we actually do is to iterate on $\sigma$ to find the 
zeros of the fitting determinant at $x_f$.

\subsection {Method of integration}

The method implemented to solve numerically the equations considered above
is a shooting method using a second-order differences representation of the equations.
It consists in writing the differential equations relating the values at two mesh points, $x_n$ and $x_{n{+}1}$, as
\begin{equation}
\vec{y}(n{+}1) = \vec{y}(n) + {h_n \over 2}\; \left[\devfd{\vec{y}}{x}(n) +
	\devfd{\vec{y}}{x}(n{+}1) \right] + {\cal O}(h_n^3)\;,
\end{equation}
where $h_n {=} x_{n{+}1} {-} x_n$ and with $\vec{y}(n) {\equiv} \vec{y}(x_n)$.
In order to replace the derivatives we use the different sets of differential equations discussed above for the regions $0{\le} x_{in} {\le} x_f {\le} x_{out} {\le} x_s$.

We also have to implement the boundary conditions.
It is done by setting the values of $y_1$ and $y_3$ at the boundaries
(centre and surface) and to calculate the values of $y_2$ and $y_4$ (at both
boundaries) from the relations constructed in the previous Section.
Both linearly independent solutions are found by setting
the central/atmospheric values of $y_1$ equal to one and $y_3$
alternatively to one and to zero.
The actual solution is a linear combination of these two (for the interior
solution - up to $x_f$, as well as for the external solution - down to $x_f$).

Since we are using a shooting method, from the values of
these two solutions (``in'' and ``out'') at $x_f$, we construct the
matching matrix
whose determinant has to be zero if $\sigma$ is an eigenvalue.
So the task of finding an eigenvalue is reduced to finding the zero of the determinant for the fitting conditions at $x_f$.

We maximize the efficiency of the search for the eigenvalues (zeros of the determinant) by using the fact that these values are separated approximately by
\begin{equation}\begin{array}{l}
\dis \Delta\sigma_p^2 \sim \left({GM \over R^3}\right)^{1/2}
	{2\pi\sigma \over \dis \int_0^R c_0^{{-}1}\diff r}
	\quad \hbox{\rm (p-modes),} \\
\dis \Delta\sigma_g^2 \sim \left({GM \over R^3}\right)^{1/2}
	{2\pi \sigma^3 \over \dis \left(l{+}{1\over 2}\right)
	\int_0^R {N_0^2 \over r} \; \diff r}
	\quad \hbox{\rm (g-modes).}
\end{array}\end{equation}

\subsection {Accuracy of the results}

The actual accuracy of the final values of the frequencies are
determined by several aspects of the calculation.
As it would be expected, the accuracy of the results (frequencies) depends on the accuracy of the equilibrium model being used.
Here we do not address this issue, referring the reader to \citet{monteiro06} and \citet{lebreton08}.

But another aspect associated with the equilibrium model,
and determining the precision of the calculated eigenvalues, is
the mesh on which the equilibrium model is given.
To minimise this effect, before calculating the frequencies we produce a re-meshing of the equilibrium model.
The actual details of the new mesh depends on the type of model and oscillation modes being calculated.
We use a receipt similar to the one discussed by \citet{cd91}.
This allows us to minimize the errors caused by having too few points where the eigenfunctions are expected to vary more strongly.

Other aspect determining the accuracy of the eigenvalues is of course
the numerical method used to integrate the four differential equations
discussed above.
In our case we have a second order scheme for the integration of the system of differential equations.
To this we have also added the use of reduced dependent variables in the regions where the amplitudes of the eigenfunctions would be otherwise very small.

Further to this the code also uses an extrapolation to improve the accuracy of the determination of each frequency.
It is known as Richardson Extrapolation.
This uses the fact that our second order integration has an error which varies with the inverse of the squared number of mesh points.
Using such a fact it can be written that the actual value of the eigenvalue is
\begin{equation}
\sigma^2 = {\alpha \over \alpha{-}1}\;\; \sigma_N^2 -
	{1 \over \alpha{-}1}\;\; \sigma_{N^\prime}^2
\quad {\rm with} \quad
	\alpha = \left({N \over N^\prime}\right)^2 \;,
\end{equation}
where $\sigma_N$ is the result found for a mesh of $N$ points and
$\sigma_{N^\prime}$ for a mesh of $N^\prime$ points.
The code uses, by default, $N^\prime{\sim}N/2$, giving $\alpha{\sim}4$.
This extrapolation requires extra work but improves significantly the accuracy of the numerical frequencies (see \citealt{moya-apss}).

\begin{figure}[ht!]
\centering
\hspace{-8pt}\includegraphics[width=\hsize]{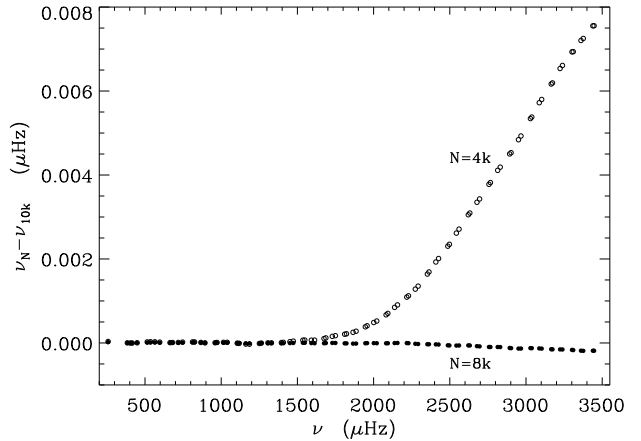}
\caption{
  Comparison of the frequencies obtained using an equilibrium model provided with a different number of mesh points.
  The reference is for a 10k mesh.
  Only differences for frequencies with $l=0,1,2,3$ and $100\muHz\le \nu \le 3500\muHz$ are shown.}
\label{fig:mesh_dif}
\end{figure}

The behaviour of the frequencies with increasing number of mesh points is an internal checkpoint that allows one to identify where the actual frequencies are no longer affected by the precision of the integration scheme. 
Such a comparison is shown in Fig.~\ref{fig:mesh_dif} where the frequencies obtained for a solar model in a mesh of 10k points is compared with the frequencies obtained using the same model with 4k and 8k mesh points.
We have also performed a detailed comparison with the results from {\sc adipls} \citep{adipls} to have an external check on the computation.
Further comparisons of \posc\ with other codes has been performed recently by \citet{moya-apss}.

In general (when using a model in a mesh of $\sim$6k points) \posc\ frequencies of oscillation for solar-type stars have an estimated numerical uncertainty below $0.001\muHz$.
This is below the current observational errors of solar frequencies.
Similar values are obtained for oscillation modes of stars of different masses and ages if the mesh is adequately adapted (in terms of number of points and its distribution) to the eigenfunctions being calculated ($g$ or $p$ modes).

\section{Output}

The main output of the code are the values of the frequencies of linear adiabatic oscillations of an equilibrium model of a star.
The calculation to be done is defined by an interval in frequency ($\omega_a\le\omega\le\omega_b$) and in mode degree ($l_a\le l\le l_b$).

\subsection {Mode classification}

The mode order of each eigenvalue is obtained using a method similar
to the phase diagram as described by \citet{unno89}.
It consists in counting the number of times the solution crosses the line $y_1{\equiv}0$ in the plane $( y_1, y_2)$.
If the cross is clockwise it counts as (-1) otherwise as (+1).
When $l{=}0$ (radial modes) an additional (+1) cross is considered.

To a total negative counting of the crosses corresponds a $g$-mode while $p$-modes have positive counting results, with the number corresponding to the mode order.
The solutions corresponding to $f$-mode eigenvalues have a total of zero counts. 

\subsection {Mode inertia and eigenfunctions}

The eigenfunctions are provided in different formats (several normalisations and/or combinations) depending on what is required.
These are obtained from $\vec{y}$ and correspond to combinations of the functions,
\begin{equation}
{\xi_r(r) \over \xi_r(R)}, \;  {\xi_h(r) \over \xi_h(R)}, \;
{\tilde{P}(r) \over \tilde{P}(R)},\; {\tilde{\Phi}(r) \over \tilde{\Phi}(R)}.
\end{equation}
The equilibrium structure is also used to calculate different normalizations of the eigenfunctions.

The code provides, in addition to the mode parameters and frequencies, the mode inertia as given by,
\begin{equation}
E_{ln} \equiv {4\pi \over M} \;
   {\dis \int_0^R \left[\xi_r^2(r) + L \; \xi_h^2(r) \right] r^2 \rho \; \diff r \over
\xi_r^2(R) + L \; \xi_h^2 (R) } \;.
\end{equation}

\section{Conclusion}

This work provides a brief description of \posc\ - the Porto Oscillation Code.
This code has been developed mainly for calculating linear adiabatic oscillations of stellar models for stars similar to the Sun (in mass).
The code is written in {\tt Fortran~77} and is modular.
It is prepared to accept input models in the {\sc amdl}\footnote{See the description of some file formats for stellar evolution models at \url{http://www.astro.up.pt/corot/ntools/}} format.
Tools are also available to convert almost any available stellar model output to the required format to be used by \posc.

The code has been applied to several cases, namely the Sun \citep{monteiro96a,monteiro96b,monteiro05} and other stars \citep{cunha03,fernandes03}, including pre-main sequence models \citep{ruoppo07}.
It has also been used to produce the frequencies of reference grids of stellar evolution models for asteroseismology \citep{marques08}.

\acknowledgements

We thank J. Christensen-Dalsgaard for all the data and documentation provided over the last 20 years that have allowed the author to implement and improve this code. 
This work was supported in part by the
European Helio- and Asteroseismology
Network (HELAS), a major international collaboration
funded by the European Commission (FP6), as well as by FCT and POCI2010 (FEDER) through projects
{\small POCI/CTE-AST/57610/2004} and {\small POCI/V.5/B0094/2005}.



\begin{thebibliography}{}

\newcommand{\thisapss}{\apss\ -- this volume} 

\bibitem[\protect\citeauthoryear{{Christensen-Dalsgaard}}{2008}]{astec}
  Christensen-Dalsgaard, J.:
{ASTEC - the Aarhus STellar Evolution Code}.
\thisapss\ (2008)

\bibitem[\protect\citeauthoryear{{Christensen-Dalsgaard}}{2008}]{adipls}
  Christensen-Dalsgaard, J.:
{ADIPLS -- the Aarhus adiabatic oscillation package}.
\thisapss\ (2008)

\bibitem[\protect\citeauthoryear{{Christensen-Dalsgaard} and {Berthomieu}}{1991}]{cd91}
  Christensen-Dalsgaard, J., Berthomieu, G.:
  {Theory of solar oscillations}.
  In Solar Interior and Atmosphere
  Cox A.N., Livingston W.C., Matthews M.S. (Eds):
  University of Arizona Press, p. 401 (1991)

\bibitem[\protect\citeauthoryear{{Cunha} {\etal}}{2003}]{cunha03}
  {{Cunha}, M.S., {Fernandes}, J.M.M.B., {Monteiro}, M.J.P.F.G.}:
  {Seismic tests of the structure of rapidly oscillating Ap stars: HR1217},
  \mnras\ 343, 831 (2003)

\bibitem[\protect\citeauthoryear{{Fernandes} and {Monteiro}}{2003}]{fernandes03}
  {{Fernandes}, J., {Monteiro}, M.J.P.F.G.}:
  {HR diagram and asteroseismic analysis of models for beta  Hydri},
  \aap\ 399, 243 (2003)

\bibitem[\protect\citeauthoryear{{Lebreton} {\etal}}{2008}]{lebreton08}
  Lebreton, Y., Monteiro, M.J.P.F.G., Montalb\'an, J., \etal:
  {The CoRoT Evolution and Seismic Tools Activity}.
  \thisapss\ (2008)

\bibitem[\protect\citeauthoryear{{Marques} {\etal}}{2008}]{marques08}
  Marques, J.P., Monteiro, M.J.P.F.G., Fernandes, J.:
{Grids of stellar mdoels and frequencies for asteorseismology ({\sc cesam + posc})}.
\thisapss\ (2008)

\bibitem[\protect\citeauthoryear{{Monteiro}}{1996}]{monteiro96a}
  Monteiro, M.J.P.F.G.:
  {Seismology of the Solar Convection Zone}.
  {PhD Thesis, Queen Mary and Westfield College University of London} (1996)

\bibitem[\protect\citeauthoryear{{Monteiro} and {Thompson}}{2005}]{monteiro05}
   {Monteiro}, M.J.P.F.G., {Thompson}, M.J.:
    {\mnras} 361, 1187 (2005)
   
\bibitem[\protect\citeauthoryear{{Monteiro} {\etal}}{1996}]{monteiro96b}
   {Monteiro}, M.J.P.F.G., {Christensen-Dalsgaard}, J., 
	{Thompson}, M.J.:
    {\aap} 307, 624 (1996)

\bibitem[\protect\citeauthoryear{{Monteiro} \etal}{2006}]{monteiro06}
  Monteiro, M.J.P.F.G.,
  Lebreton, Y.,
  Montalb\'an, J.,
  \etal:
{Report on the CoRoT Evolution and Seismic Tools Activity}.
In: The CoRoT Mission, 
  M. Fridlund, A. Baglin, J. Lochard {\&} L. Conroy (Eds):
  ESA SP-1306, 363 (2006)


\bibitem[\protect\citeauthoryear{{Moya} \etal}{2008}]{moya-apss}
  Moya,~A., 
  Christensen-Dalsgaard,~J.,
  Charpinet,~S.,
  \etal:
\thisapss\ (2008)

\bibitem[\protect\citeauthoryear{{Ruoppo} {\etal}}{2007}]{ruoppo07}
   {{Ruoppo}, A., {Marconi}, M., {Marques}, J.P., \etal}:
   {A theoretical approach for the interpretation of pulsating PMS 
   intermediate-mass stars},
   \aap\ 466, 261 (2007)

\bibitem[\protect\citeauthoryear{{Unno} {\etal}}{1989}]{unno89}
  {Unno}, W., {Osaki}, Y., {Ando}, H., {Saio}, H., {Shibahashi}, H.:
    {Nonradial oscillations of stars} (2nd edition).
    University of Tokyo Press (1989)

\end{thebibliography}
\end{document}